\documentstyle[12pt]{article}

\normalbaselines
\font\tenbm=cmmib10
\font\sevenbm=cmmib7
\font\fivebm=cmmib5
\newfam\bmfam
\textfont\bmfam=\tenbm \scriptfont\bmfam=\sevenbm
\scriptscriptfont\bmfam=\fivebm
{\count0=\number\bmfam \multiply\count0 by "100
\def\defbgreek#1#2#3{{\count1=\count0 \advance\count1 by "#2#3
  \global\mathchardef#1=\count1 }}
\defbgreek\balpha  0B \defbgreek\brho       1A
\defbgreek\bbeta   0C \defbgreek\bsigma     1B
\defbgreek\bgamma  0D \defbgreek\btau       1C
\defbgreek\bdelta  0E \defbgreek\bupsilon   1D
\defbgreek\bepsilon0F \defbgreek\bphi       1E
\defbgreek\bzeta   10 \defbgreek\bchi       1F
\defbgreek\bmeta   11 \defbgreek\bpsi       20
\defbgreek\btheta  12 \defbgreek\bomega     21
\defbgreek\biota   13 \defbgreek\bvarepsilon22
\defbgreek\bkappa  14 \defbgreek\bvartheta  23
\defbgreek\blambda 15 \defbgreek\bvarpi     24
\defbgreek\bmu     16 \defbgreek\bvarrho    25
\defbgreek\bnu     17 \defbgreek\bvarsigma  26
        \defbgreek\bxi     18 \defbgreek\bvarphi    27
\defbgreek\bpi     19}

\input{tcilatex}

\begin{document}

\author{Yuri A.Rylov}
\title{Specification of statistical description in quantum mechanics}
\date{Institute for Problems in Mechanics, Russian Academy of Sciences \\
101-1 ,Vernadskii Ave., Moscow, 117526, Russia \\
email: rylov@ipmnet.ru\\
Web site: {$http://rsfq1.physics.sunysb.edu/\symbol{126}rylov/yrylov.htm$}\\
or mirror Web site: {$http://194.190.131.172/\symbol{126}rylov/yrylov.htm$}}
\maketitle

\begin{abstract}
It is shown that the statistical conception of quantum mechanics is
dynamical but not probabilistic, i.e. the statistical description in quantum
mechanics is founded on dynamics. A use of the probability theory, when it
takes place, is auxiliary. Attention is drawn to the fact that in the
quantum mechanics there are two different objects: an individual object to
be statistically described and a statistical average object, which is a
result of the statistical description. Identification of the two different
objects (a use of the same term for both) is an origin of many known quantum
mechanics paradoxes.
\end{abstract}

Without question the quantum mechanics is a statistical theory, because its
mathematical formalism works with quantities distributions (wave functions),
and many of its results are formulated in terms of the probability theory.
But on the outside the quantum mechanics appears as a dynamical conception.
In the end of XIXth century one succeeded to substantiate the thermodynamics
as a statistical description of chaotically moving molecules, and many
investigators believed that the same is true for quantum mechanics. But
attempts \cite{M49,F52} of the quantum mechanics formulation in terms of the
probability theory failed. Unfortunately, trying to represent the quantum
mechanics as a statistical description of random particle motion, one
overlooks usually, that the random component of the particle motion may be
relativistic, whereas the regular component remains nonrelativistic.

The probability theory, employed successfully in the statistical physics for
statistical description of chaotic molecules motion, is not suitable for
description of random relativistic motion. The fact is that an application
of the concept the probability density supposes that the set of all possible
system states can be separated into sets of independent simultaneous events.
It is impossible for relativistic continuous dynamic system, because
absolute simultaneity is absent in the relativity theory. One cannot use
simultaneity in some coordinate system, because any coordinate system is a
way of description. A joint employment of the probability theory and of the
conditional simultaneity (simultaneity in some coordinate system) means a
statistical description of the description methods, whereas one needs a
statistical description of states of the considered dynamic system.

Refusing the probability theory employment, one can overcome the originated
difficulties. It should note here, that successful employment of the
probability theory in the statistical physics leads to a confusion, when
some physicists consider terms ''statistical description'' and
''probabilistic description'' to be synonyms. They cannot imagine that there
is a statistical description without a use of the probability theory. But
the term ''statistical description'' means only, that one considers many
similar or almost similar objects. The probability theory may be used
optionally at the statistical description. The term ''probabilistic
description'' means an employment of the probability theory, which imposes a
set of constraints on the way of description. For instance, the probability
density is to be nonnegative that is succeeded to be fulfilled not always.

In the nonrelativistic physics the physical object that should be described
statistically is a particle, i.e. a point in the usual space or in the phase
one. The density of points in the space is nonnegative, and this fact can be
a ground for introduction of the probability density concept. In the
relativistic theory the physical object that should be described
statistically is a world line in the space-time. The density of world lines
in the vicinity of some point $x$ of the space-time is a 4-vector, which
cannot be a ground for introduction of the probability density concept. The
alternative version, when any world line is considered to be a point in some
space ${\cal V}$, admits one to introduce a concept of the world line
probability density in the space ${\cal V}$ of world lines. But such a
description is nonlocal, because two world lines, coinciding everywhere
except for some remote region, are represented by two different points in $%
{\cal V}$. In general, these points will not be close. In other words, such
an introduction of the probability density is very unsuccessful.

The way out is a refuse of the probability theory employment at a
statistical description. The dynamical conception of statistical description
(DCSD) is used instead of the probabilistic conception. Instead of the
stochastic system ${\cal S}_{{\rm st}}$, for which there are no dynamic
equations, one uses the set ${\cal E}[N,{\cal S}_{{\rm st}}]$, consisting of
large number $N$ of similar independent systems ${\cal S}_{{\rm st}}$ and
called statistical ensemble of systems ${\cal S}_{{\rm st}}$. The
statistical ensemble ${\cal E}[N,{\cal S}_{{\rm st}}]$ forms a deterministic
dynamic system. There are dynamic equations for this system, although there
are no dynamic equations for elements ${\cal S}_{{\rm st}}$ of the
statistical ensemble. The statistical description consists in investigation
of the ensemble ${\cal E}[N,{\cal S}_{{\rm st}}]$ properties as a
deterministic dynamic system. On the basis of this investigation one draws a
conclusion on properties of the statistical ensemble elements (stochastic
systems ${\cal S}_{{\rm st}}$). The concept of probability may be not used,
as far as one investigates a dynamic system and its properties.

DCSD is not so informative as the probabilistic statistical description in
the sense that some conclusions and estimations, which could be made at the
probabilistic description, cannot be derived in the scope of DCSD. But we
must take this, because a more informative statistical description is
impossible. The circumstance that one perceives the quantum mechanics as a
dynamical (but not as a statistical, i.e. probabilistic) conception is
connected with a use of DCSD, that in turn is conditioned by ''relativistic
roots'' of nonrelativistic quantum mechanics. Note that DCSD is an universal
conception in the sense that it may be used in both relativistic and
nonrelativistic cases.

The lesser (as compared with the probabilistic description) informativity of
description in the scope of DSCD is displayed, for instance, in the fact
that a simultaneous distribution over positions and momenta is absent at a
pure state. Moreover distribution $|\langle p|\psi \rangle |^{2}$ over the
particle momenta at the pure state, described by the wave function $|\psi
\rangle $, has a formal character, and it cannot be tested experimentally.
Of course, one can obtain the distribution $|\langle p|\psi \rangle |^{2}$
experimentally. It is sufficient to drop a flux of particles onto
diffraction grating and to investigate the obtained diffraction picture. But
the quantum mechanics technique supposes, that obtained in such a way
distribution over momenta must be attributed to some state (some wave
function $|\psi \rangle $). Derivation of the distribution $|\langle p|\psi
\rangle |^{2}$ needs a long time. This time is the longer the more exact
distribution is to be obtained. In this time the wave function changes, and
it is not clear to what wave function the obtained distribution over momenta
should be attributed. The performed analysis \cite{R77} shows that the
momentum distribution, obtained experimentally, cannot be attributed to any
state (wave function). It means that the distribution over momenta is
formal, i.e. it cannot be tested experimentally.

Along with the statistical ensemble ${\cal E}[N,{\cal S}]$ of systems ${\cal %
S}$, or even instead of it one can introduce the statistical average dynamic
system $\left\langle {\cal S}\right\rangle $, which is defined as the
statistical ensemble ${\cal E}\left[ N,{\cal S}\right] $, ($N\rightarrow
\infty $), normalized onto one system. Mathematically it means that, if $%
{\cal A}_{{\rm E}}\left[ N,d_{N}\left\{ X\right\} \right] $ is the action
for ${\cal E}\left[ N,{\cal S}\right] $, then
\[
\left\langle {\cal S}\right\rangle :\;\;{\cal A}_{\left\langle
S\right\rangle }\left[ d\left\{ X\right\} \right] =\lim_{N\rightarrow \infty
}\frac{1}{N}{\cal A}_{{\rm E}}\left[ N,d_{N}\left\{ X\right\} \right]
,\qquad d\left\{ X\right\} =\lim_{N\rightarrow \infty }d_{N}\left\{
X\right\}
\]
is the action for $\left\langle {\cal S}\right\rangle $, where $X$ is a
state of a single system ${\cal S}$, and $d_{N}\left\{ X\right\} $ is a
distribution, describing in the limit $N\rightarrow \infty $ both the state
of the statistical ensemble ${\cal E}\left[ N,{\cal S}\right] $, and the
state of the statistical average system $\left\langle {\cal S}\right\rangle $%
. In particular, $d\left\{ X\right\} $ may be a wave function.

Substitution of the statistical ensemble ${\cal E}\left[ N,{\cal S}\right] $
by the statistical average system $\left\langle {\cal S}\right\rangle $ is
founded on insensibility of the statistical ensemble to the number $N$ of
its elements, provided $N$ is large enough. The statistical average system $%
\left\langle {\cal S}\right\rangle $ is a kind of the statistical ensemble.
Formally it follows from the fact, that the state of $\left\langle {\cal S}%
\right\rangle $, as well as the state of the statistical ensemble ${\cal E}%
\left[ N,{\cal S}\right] $, is described by the distribution $d_{N}\left\{
X\right\} $, $N\rightarrow \infty $, whereas the state of a single system $%
{\cal S}$ is described by the quantities $X$, but not by their distribution.
Using this formal criterion, one can distinguish easily between the single
dynamic system ${\cal S}$ and the statistically averaged system $%
\left\langle {\cal S}\right\rangle $. For instance, if the state of a system
is described by the wave function $|\psi \rangle $, i.e. by some complex
distribution, it is clear that one deals with a statistical average system $%
\left\langle {\cal S}\right\rangle $, and any statement that the wave
function describes some state of individual particle ${\cal S}$ (even
quantum one) are not justified.

Indeed, let ${\cal A}_{{\rm S}}\left[ \psi \right] $ be the action,
describing  ''quantum particle'', and the Schr\"{o}din\-ger equation be
dynamic equation generated by this action. What is the ''quantum particle''?
Is it a single particle ${\cal S}_{{\rm S}}$, or a statistical average
particle $\left\langle {\cal S}_{{\rm S}}\right\rangle $? Transforming the
action ${\cal A}_{{\rm S}}\left[ \psi \right] $ to the form, where the
4-current
\[
j^{i}=\left\{ \rho ,{\bf j}\right\} ,\qquad \rho =\psi ^{\ast }\psi ,\qquad
{\bf j}=-\frac{i\hbar }{2m}\left( \psi ^{\ast }{\bf \nabla }\psi -{\bf %
\nabla }\psi ^{\ast }\cdot \psi \right)
\]
is one of dependent dynamic variables \cite{R95} and tending $\hbar
\rightarrow 0$, one obtains the action ${\cal A}_{{\rm Scl}}\left[ \rho ,%
{\bf j},\varphi \right] $, describing a statistical ensemble ${\cal E}\left[
1,{\cal S}_{{\rm cl}}\right] =\left\langle {\cal S}_{{\rm cl}}\right\rangle ,
$ where ${\cal S}_{{\rm cl}}$ is the classical particle, described by the
Hamiltonian $H={\bf p}^{2}/2m$. If the ''quantum particle'' be a single
dynamic system ${\cal S}_{{\rm S}}$, one would obtain in the limit $\hbar
\rightarrow 0$ the action for a single particle ${\cal S}_{{\rm cl}}$ (but
not for $\left\langle {\cal S}_{{\rm cl}}\right\rangle $). This example
shows that there is no foundation for considering the ''quantum particle''
as an individual particle.

As far as the quantum mechanics is a statistical conception, it contains two
different objects: individual object and statistical average object, which
should be distinguished and not be confused between themselves. In this
paper we are going to show that many paradoxes of quantum mechanics
(Schr\"{o}dinger cat, reduction of the wave packet under measurement) do not
appear, if the statistical character of quantum mechanics is taken into
account consequently.

Any statistical description is a description of many identical or similar
objects ${\cal S}$. The statistical average object $\left\langle {\cal S}%
\right\rangle $ is a result of the statistical description. If any
individual object ${\cal S}$ is described by quantities $X$, then the
statistical average object $\left\langle {\cal S}\right\rangle $ is
described by a distribution of these quantities. At an approximate
description the statistical average object $\left\langle {\cal S}%
\right\rangle $ may be described by average values $\left\langle
X\right\rangle $ of quantities $X$.

For instance, the statistical description of Moscow inhabitants introduces a
conception of a statistical average inhabitant. For the statistical average
inhabitant one can calculate the mean age, the mean weight, the mean height
etc. But the statistical average inhabitant can be attributed by a
distribution over age, a distribution over weight, a distribution over
height etc. It is possible further complication of statistical description,
for instance, a consideration of correlation between the age and the height,
between the sex and the height etc.. The statistical average inhabitant of
Moscow is half-man -- half-woman. Of course, nobody of inhabitants of the
Earth concludes that Moscow is peopled by hermaphrodites. It means simply
that one half of inhabitants cosists of men and other half of them consists
of women. However, wise inhabitants of some remote planet, which did not see
Moscow inhabitants, can conclude on the basis of this information that
Moscow is inhabited by hermaphrodites. Such a conclusion will not be
surprising, because different interpretation of such a statistical
information is connected with the different knowledge on individual
inhabitants of Moscow. Inhabitants of the Earth know that hermaphrodites are
very seldom among peoples, but inhabitants of other planet may not know
this. The problem of interpretation of statistical characteristics can
appear to be complicated, if information on individual objects of
statistical description is absent or not complete enough.

Experiments show that the microparticle motion is stochastic. Experiments
with a single electron ${\cal S}$ are not reproducable, in general, whereas
distributions of results, obtained in experiments with many single electrons
${\cal S}$, are reproduced at repeating of experiments. Hence, if
investigating an electron, we want to have deterministic dynamic equations,
describing evolution of its state, the electron state is to be described by
some distribution $d\{X\}$ of its characteristics $X$. In other words,
dynamic equations should be written for statistical average electron $%
\left\langle {\cal S}\right\rangle $.

The object of quantum mechanical investigation is a statistical average
particle $\left\langle {\cal S}\right\rangle $. Its state is described by
the wave function, i.e. by some complex distribution $d\{X\}$ of its
characteristics $X$. Dynamic equation (Schr\"{o}dinger equation) describes
evolution of the wave function. In other words, quantum mechanics technique
deals only with the abstract statistical average particle. What is interplay
between the abstract statistical average particle $\left\langle {\cal S}%
\right\rangle $ and the real particle ${\cal S}$, it is a special problem.
But this problem does not appear at the work with the quantum mechanics
technique, as far as this technique does not deal with single particles.

Problem of interrelation between individual particle ${\cal S}$ and
statistical average particle $\left\langle {\cal S}\right\rangle $ appears
only at a consideration of the measurement process. The measurement with an
individual particle ${\cal S}$ (it will be referred to as $S$-measurement)
leads always to a definite result, i.e. at $S$-measurement one obtains some
number $R^{\prime }$, giving the value of the quantity ${\cal R}$, measured
at this experiment. The question whether the wave function $\psi $ of this
particle ${\cal S}$ changes at this experiment, has no sense at all, as far
as the particle ${\cal S}$ is not described by the wave function. The wave
function $\psi $ describes the state of the statistical average particle $%
\left\langle {\cal S}\right\rangle $, and it is an attribute of the particle
$\left\langle {\cal S}\right\rangle $.

Is it possible to carry out measurement with the main object of quantum
mechanics -- the statistical average particle $\left\langle {\cal S}%
\right\rangle $? Yes, it is possible, but it is another measurement (but not
$S$-measurement). Measurement with statistical average particle $%
\left\langle {\cal S}\right\rangle $ is a mass measurement ($M$%
-measurement), i.e. a set of many $S$-measurements). one carries out the $M$%
-measurement with the statistical ensemble ${\cal E}\left[ N,{\cal S}\right]
$, consisting of $N$, ($N\rightarrow \infty $) particles ${\cal S}$,
prepared by the same method. On one hand, the statistical ensemble ${\cal E}%
\left[ \infty ,{\cal S}\right] $ is a dynamic system, whose state is
described by the distribution $d\{X\}$ of characteristics $X$ of particles $%
{\cal S}$. On the other hand, one can consider the same statistical ensemble
${\cal E}\left[ N,{\cal S}\right] $, ($N\rightarrow \infty $) to be
consisted of $N$ similar statistical average particles $\left\langle {\cal S}%
\right\rangle $ at the state, described by the distribution $d\{X\}$ (in the
simplest case it is a wave function $\psi $), and, hence, its state is
described by the same distribution $d\{X\}$ (or by the same wave function $%
\psi $), as $\left\langle {\cal S}\right\rangle $.

Producing a set of $N$ $S$-measurements, one obtains at any measurement, in
general, different values $R^{\prime }$ of measured quantity ${\cal R}$.
Statements of quantum mechanics on influence of the measurement on the state
of the measured particle $\left\langle {\cal S}\right\rangle $ are
statements on properties of $M$-measurement. Carrying out an $M$-measurement
upon a statistical average particle $\left\langle {\cal S}\right\rangle $,
its state turns from a pure state, described by the wave function $\psi $,
to, in general, a mixed state, described by the density matrix $\rho $. Why
and how does it take place?

According to quantum principles the action of a measuring device ${\cal M}$
on the measured statistical average particle $\left\langle {\cal S}%
\right\rangle $ may be only a change of the Hamiltonian $H$, describing
evolution of statistical average system $\left\langle {\cal S}\right\rangle $.
 However, no matter what this change of Hamiltonian may be, the state
evolution happens to be a such one that the pure state $\psi $ of
statistical average system $\left\langle {\cal S}\right\rangle $ remains to
be pure. Nevertheless, action of the measuring device ${\cal M}$ on $%
\left\langle {\cal S}\right\rangle $ at the state $\psi $ leads to a passage
of $\left\langle {\cal S}\right\rangle $ into a mixed state $\rho $.
Qualitatively it is explained by that the measuring device ${\cal M}$
transforms the total Hamiltonian $H$ in several different Hamiltonians $%
H_{1} $, $H_{2},\ldots $, each of them depends on the state $\varphi _{k}$, $%
k=1,2,\ldots $ of measuring device ${\cal M}$.

Let us consider those particles ${\cal S}$ of the statistical ensemble $%
{\cal E}\left[ N,{\cal S}\right] $, which gave the value $R_{1}$ at the
measurement of the quantity ${\cal R}$. They form subensemble ${\cal E}_{1}%
\left[ N_{1},{\cal S}\right] $ of the statistical ensemble ${\cal E}\left[ N,%
{\cal S}\right] $. At the fixed value $R_{1}$ of the measured quantity $%
{\cal R}$ the measuring device ${\cal M}$ is found at the state $\varphi
_{1} $. Then subensemble ${\cal E}_{1}\left[ N_{1},{\cal S}\right] $ of
particles ${\cal S}$, which gave the result $R_{1}$ at the measurement, will
evolve with the Hamiltonian $H_{1}$. The subensemble ${\cal E}_{2}\left[
N_{2},{\cal S}\right] $ of particles ${\cal S}$, which gave the result $%
R_{2} $ at the measurement, will evolve with the Hamiltonian $H_{2}$,
because the measuring device ${\cal M}$ is found now in another state $%
\varphi _{2}$. Each value $R_{k}$ of the measured quantity ${\cal R}$
associates with an evolution of the subensemble ${\cal E}_{k}\left[ N_{k},%
{\cal S}\right] $ of the statistical ensemble ${\cal E}\left[ N,{\cal S}%
\right] $ with the Hamiltonian $H_{k}$.

The number $N_{k}$ of particles in the corresponding subensemble ${\cal E}%
_{k}\left[ N_{k},{\cal S}\right] $ is proportional to the probability of the
measured value $R_{k}$ of the measured quantity ${\cal R}$. Evolution of
different subensembles ${\cal E}_{k}\left[ N_{k},{\cal S}\right] $ is
different. As a corollary one cannot speak about one wave function,
describing the state of the whole ensemble ${\cal E}\left[ N,{\cal S}\right]
$. One should speak about states of subensembles ${\cal E}_{k}\left[ N_{k},%
{\cal S}\right] $, constituting the statistical ensemble ${\cal E}\left[ N,%
{\cal S}\right] $.

In general, any $N$-measurement may be conceived as an abstract single
procedure, produced on the statistical average system $\left\langle {\cal S}%
\right\rangle $. Action of $M$-measurement on statistical average system $%
\left\langle {\cal S}\right\rangle $ is described formally by the rule of
von Neumann \cite{N32}. For the reduction process it is important that
result $R_{k}$ of measurement of the quantity ${\cal R}$ be fixed, i.e. that
the measuring device ${\cal M}$ be found at the state $\varphi _{k}$,
because only in this case one can speak on a definite Hamiltonian $H_{k}$,
which determines evolution of ${\cal E}_{k}\left[ N_{k},{\cal S}\right] $.
It is this procedure of measurement ($M$-measurement) that is considered in
most of papers \cite{N32,M98}.

Although for simplicity the statistical ensemble ${\cal E}\left[ \infty ,%
{\cal S}\right] $, consisting of many independent stochastic similar system $%
{\cal S}$, may be conceived as a more simple statistical average system $%
\left\langle {\cal S}\right\rangle $, one should remember, that the state of
both systems ${\cal E}\left[ \infty ,{\cal S}\right] $ and $\left\langle
{\cal S}\right\rangle $ is described by the same distribution $d\{X\}$.
Similarly, although the $M$-measurement, consisting of many $S$%
-measurements, may be conceived as a single measurement procedure, acting on
the state of the statistical average system $\left\langle {\cal S}%
\right\rangle $ according to the rule of von Neumann, nevertheless one
should remember the origin of this rule and not to identify $M$-measurement
with $S$-measurement, if for no reason than they are applied to different
systems (respectively to $\left\langle {\cal S}\right\rangle $ and to ${\cal %
S}$), whose identification is not admissible. In all doubtful cases one
should return to the concept of $M$-measurement as a procedure, consisting
of many independent $S$-measurements. The statistical average system $%
\left\langle {\cal S}\right\rangle $ should be conceived as a statistical
ensemble normalized on one system, as far as the statistical ensemble ${\cal %
E}\left[ \infty ,{\cal S}\right] $ cannot be confused with an individual
system ${\cal S}$ (but, the experience suggests, it is possible for $%
\left\langle {\cal S}\right\rangle $ and ${\cal S}$).

Is it possible to speak on derivation of a definite value $R_{k}$ at $M$%
-measurement of the quantity ${\cal R}$ for statistical average particle $%
\left\langle {\cal S}\right\rangle $? It is possible, but it is a new kind
of measurement, so-called selective $M$-measurement, or $SM$-measurement. $%
SM $-measurement is $M$-measurement, accompanied by a selection of only
those systems ${\cal S}$, for which a single $S$-measurement gives the same
measurement result $R_{k}$. It is of no importance, who or what carries out
this selection. This may be device, human being, or environment. The
selection is introduced directly in the definition of the $SM$-measurement,
and it is to be made by anybody. The process of selection of individual
particles ${\cal S}$ may be interpreted as a statistical influence of the
measuring device ${\cal M}$ on the statistical average particle $%
\left\langle {\cal S}\right\rangle $.

The statistical influence in itself is not a force interaction. It is an
influence of the measuring device ${\cal M}$, leading to a selection of some
and discrimination of other elements of the statistical ensemble ${\cal E}%
\left[ \infty ,{\cal S}\right] $. In general one may say on statistical
influence of the measuring device ${\cal M}$ on the state of the measured
statistical average system $\left\langle {\cal S}\right\rangle $. The state
of the system $\left\langle {\cal S}\right\rangle $ is determined by the
distribution $d\{X\}$ of quantities $X$, describing the state of the system $%
{\cal {S}}$. Statistical influence of the measuring device ${\cal M}$ on $%
\left\langle {\cal S}\right\rangle $ leads to a change of this distribution $%
d\{X\}$.

It is worth to note that the quantum mechanics makes predictions, concerning
only $M$-measurements. Any quantum mechanical prediction can be tested only
by means of $M$-measurement. There is no quantum mechanical predictions that
could be tested by means of one single measurement ($S$-measurement). Result
of individual measurement ($S$-measurement) of the quantity ${\cal R}$ can
be predicted never. All that can be made, using quantum mechanical
technique, is a probability of the fact that at the $S$-measurement of the
quantity ${\cal R}$ the result $R^{\prime}$ will be obtained. But a
prediction of the result probability does not mean a prediction of the
result in itself.

To understand what means the prediction of the result probability from the
measurement viewpoint, let us consider the following situation. Let a
calculation on the basis of quantum mechanics technique gives, that the
probability of obtaining the result $R^{\prime }$ at the measurement of the
quantity ${\cal R}$ in the system at the state $|\psi \rangle $ is equal to
1/2. How can one test that this probability is equal to 1/2, but not, for
instance, to 3/4 or 1/4? It is clear that it is impossible at one single
measurement of the quantity ${\cal R}$. To test the prediction, it is
necessary to carry out $N$, ($N\rightarrow \infty $) individual
measurements, and the part of measurements, where the value $R^{\prime }$ of
the quantity ${\cal R}$ is obtained, gives the value of probability. It is
valid even in the case, when predicted probability is equal to unity. In
this case for the test of the prediction an individual experiment is also
insufficient. To test the prediction, one needs to carry out a set of many
individual measurements of the quantity ${\cal R}$, and the prediction will
be valid, provided the value $R^{\prime }$ is obtained in all cases.

At this point the quantum mechanics distinguishes from the classical
mechanics, where results of repeated measurements coincide. The classical
mechanics accepts that two different individual measurements, produced on
the system at the same state give similar results. The classical mechanics
supposes that it is possible one not to test this circumstance, and it
predicts the value $R^{\prime }$ of the measured quantity ${\cal R}$
(probability of the value $R^{\prime }$ is accepted to be equal to 1 and is
not tested).

The quantum mechanics admits that two different individual measurements,
produced on the systems at the same state, may give different results, and
quantum mechanics predicts only probability of the value $R^{\prime }$ of
the measured quantity ${\cal R}$ (but not the value $R^{\prime }$ itself).
At the test only value of the probability is verified (it is tested, even if
this value is equal to 1 or to 0). For such a test one needs a $M$%
-measurement. In other words, two predictions: (1) measurement of the
quantity ${\cal R}$ must give the value $R^{\prime }$ and (2) measurement of
the quantity ${\cal R}$ must give the value $R^{\prime }$ with the
probability 1 are two different predictions, tested by measurements of
different kinds. Predictions of the first type can be made only by classical
mechanics. The quantum mechanics can make only prediction of the second type.

All this means that the quantum mechanical technique and its predictions
deal only with mass measurements ($M$-measurement) and have nothing to do
with individual measurements. In general, appearance of the term
''probability'' in all predictions of quantum mechanics is connected with
the fact that the quantum mechanical technique deals only with distributions
$d\{X\}$ of quantities $X$, which are reproduced at repeated measurements,
but not with the quantities $X$ themselves, whose values are random. It
means that the quantum mechanics technique deals with statistical average
objects $\langle {\cal S}\rangle $ (or with statistical ensembles of single
systems).

Let us consider now well known paradox of ''Schr\"{o}dinger cat''. At first,
let us present it in the conventional manner. There is a cat in a closed
chamber. The cat's life is determined by the state of a radioactive atom,
placed in this chamber. While the atom is not decayed the cat is alive. As
soon as the atom decays, the cat becomes dead. The state of the atom is a
linear superposition of states of the nondecomposed atom and decomposed
atom. Respectively the state of the cat is a linear superposition of the
alive cat and of the dead one. Paradox consists in the simultaneous
existence of the dead cat and of the alive cat. If one opens the chamber and
observes the cat, the cat passes instantly from the state, where the cat is
neither dead, nor alive, to the definite state, where the cat is either
dead, or alive.

This paradox is a result of simple misunderstanding, when one identifies two
different object: the real individual Cat and an abstract statistical
average $\langle \mbox{cat}\rangle $. The wave function describes the state
of the abstract statistical average $\langle \mbox{atom}\rangle $ and the
state of the abstract statistical average $\langle \mbox{atom}\rangle $
determines the state of the statistical average $\langle \mbox{cat}\rangle $%
. The wave function has nothing to do with the real Cat in the chamber. The
statistical average $\langle \mbox{cat}\rangle $ bears on the real Cat the
same relation as the statistical average inhabitant of Moscow bears on a
real Ivan Sidorov, living somewhere in Leninsky avenue. If we produce $S$%
-observation, i.e. we open one definite chamber and found there an alive
cat, we have no reasons for the statement, that opening the chamber, we
change the state of $\langle \mbox{cat}\rangle $. No paradox appears at the
opening of the chamber. If we produce $M$-observation, i.e. if we consider $%
N $, ($N\gg 1$) chambers with cats, then opening them all simultaneously, we
do not discover a definite result. In some chambers one discovers alive
cats, in other ones the cats are dead. In this case observation of the state
of the abstract statistical average $\langle ${\rm cat}$\rangle $ leads to a
change of the state in the sense that the state turns from pure to mixed.
But there is no paradox.

Finally, if we carry out a selective mass measurement ($SM$-measurement,
i.e. one opens simultaneously $N$, ($N\gg 1$) chambers and chooses those of
them, where there are alive cats, then, on one hand, one obtains a definite
result (alive cats), but on the other hand, there is many alive cats and
they form a statistical ensemble, whose state is described by a certain
definite wave function. But any paradox does not appears, as far as the
reason of a change of the wave function of statistical average $\langle %
\mbox{кота}\rangle $ is evident. It is a selection of alive cats from the
total set, consisting of alive and dead cats.

The Schr\"{d}inger cat paradox is a special case of the paradox, which is
discerned sometimes in the wave function reduction, appearing as a result of
a measurement. To avoid this paradox, it is sufficient to follow a simple
logic rule, which asserts: '' One may not use the same term for notation of
different objects. (But it is admissible to use several different terms for
notation of the same object.)''. In the given case this rule is violated.
One uses the same term for the individual object and for the statistical
average object. As a corollary one uses the same term for two different
measurement processes. Note that the formal representation of $M$%
-measurement as a single act of influence on the statistical average system $%
\left\langle {\cal S}\right\rangle $ makes a large contribution into
identification of $S$-measurement with $M$-measurement. This representation
removes distinction between the $S$-measurement and the $M$-measurement and
gives the impression that $S$-measurement and the $M$-measurement are
identical procedures. After such an identification the measurement process
would begin to take on contradictory properties. On one hand, such a
measurement leads to a definite result ($S$-measurement), on the other hand,
its result is a distribution of the measured quantity ($M$-measurement). On
one hand, the measurement ($S$-measurement) is a single act, produced on $%
{\cal S}$, and the measurement does not bear on the wave function, and on
the statistical average system $\left\langle {\cal S}\right\rangle $. On the
other hand, the measurement ($M$-measurement) leads to a reduction of the
state of the system $\left\langle {\cal S}\right\rangle $.

In general, violation of the formal logic rule cited above is only a premise
for appearance of paradoxes and contradictions, but this violation does not
lead to paradoxes with a necessity. But only a few of investigators, using a
representation on $M$-measurement as a single act, succeeded to avoid
identification of $S$-measurement and $M$-measurement and inconsistencies
following from this identification. One of these few investigators was J.
von Neumann. In his monograph \cite{N32} he employed the same term for
individual and statistical average objects. Nevertheless, he obtains never
inconsistencies and paradoxes. However, ordinary investigators, who has not
powerful intellect of von Neumann, are recommended to follow the simple
logic rule cited above.

As soon as one follows this simple rule, many exotic interpretation are
removed by themselves. For instance, the Everett -- Weeler interpretation
\cite{E57,DG73}, founded on the erroneous conception, that a measurement
with an individual system ($S$-measurement) changes its wave function.

It should note that sometimes it is difficult to determine whether some
property is a property of an individual object or that of a statistical
average one. For instance, whether the half-integer spin of an electron is a
property of a single electron or a property of statistical average electron
(statistical ensemble). At the conventional approach, when one does not
distinguish between an individual electron and a statistical average
electron, such a problem does not exist. At a more careful approach we are
forced to state that properties of an individual electron have been
investigated very slightly. In most of cases we deal with measurements,
produced on many electrons, as far as only such experiments are
reproducible. As a corollary many conclusions on properties of a single
electron are unreliable.

For instance, it is a common practice to consider the Stern -- Gerlach
experiment (when the electron beam, passing the region with inhomogeneous
magnetic field, splits into two beams) to be an evidence of the statement
that each single electron has a definite spin and corresponding projection
of magnetic moment onto the magnetic field direction. In reality, the Stern
-- Gerlach experiment shows only that the Hamiltonian, describing a motion
of statistical average electron, has two discrete eigenstates, labelled by
the magnetic quantum number and distinguishing by their energy. The
question, whether discreteness of the magnetic quantum number is connected
with a discreteness in properties of a single electron, remains open,
because a discreteness of energy levels is not connected directly with with
a discrete character of interaction. For instance, the energy levels of a
spinless charged particle in a Coulomb electric field are discrete, although
there is no discreteness in the properties of a single particle. It is
doubtless that the beam split is connected with ''the electron magnetic
moment'', because it is proportional to the magnetic field gradient. But the
question remains open, whether the ''electron magnetic moment' is a property
of the individual electron ${\cal S}$ or a collective property of the
statistical average electron $\left\langle {\cal S}\right\rangle $ (see for
details \cite{R95}). Note that at identification of $\left\langle {\cal S}%
\right\rangle $ and ${\cal S}$ the problem does not appear.

Sometimes one considers, that the spectrum of electromagnetic radiation,
emanated by an excited atom is a property of individual (but not statistical
average) atom. It is adduced the argument that the modern technique admits
to confine a single atom in a trap. Then one can investigate its energy
levels and spectrum of radiation. But one overlooks that the spectrum of the
atom radiation cannot be measured as a result of a single measurement ($S$%
-measurement). The atom radiation spectrum is obtained as a result of many
measurements of radiation of the same atom, whose state is prepared by the
same way, i.e. that essentially one measures radiation of a statistical
average atom. To carry out such a $M$-measurement, it is of no importance,
whether one produces one measurement with many similarly prepared atoms, or
one produces many measurements with a single atom, preparing its state many
times by the same way. In both case one deals with the statistical average
atom.

Finally, we should like to express our astonishment in connection with a
surprising situation. How might one not to recognize during the course of 75
years of the quantum mechanics existence that one may not identify the
statistical average particle $\left\langle {\cal S}\right\rangle $,
described by the quantum mechanics technique, with the single real particle $%
{\cal S}$, existing in nature? It was being well known for a long time, that
the quantum mechanics is a statistical conception.

\end{document}